\documentclass{sf2a-conf2013}
\usepackage{graphicx}
\usepackage{hyperref}
\usepackage[]{natbib}  
\usepackage{epstopdf}

\def\BibTeX{{\rm B\kern-.05em{\sc i\kern-.025em b}\kern-.08em
    T\kern-.1667em\lower.7ex\hbox{E}\kern-.125emX}}
\bibpunct{(}{)}{;}{a}{}{,}  

\begin{document}

\TitreGlobal{SF2A 2013}


\title{Gaia, counting down to launch}

\runningtitle{Gaia}

\author{Anthony G.A. Brown}\address{Sterrewacht Leiden, Leiden University, P.O.\ Box 9513, 2300 RA Leiden,
The Netherlands}

\setcounter{page}{237}

\maketitle

\begin{abstract}
  In this contribution I provide an overview of the the European Space Agency's Gaia mission just
  ahead of its launch scheduled for November 2013.
\end{abstract}


\section{The Gaia mission}

\label{sec:Gaia}

ESA's Gaia mission is the next European breakthrough in astrophysics, a cornerstone mission
scheduled for launch in November 2013 which is designed to produce the most accurate 3D map of the
Milky Way to date \citep{Perryman2001}. The scientific power of Gaia rests on the combination of
three desirable qualities in a single mission: (1) the ability to make very accurate (global and
absolute) astrometric measurements; (2) the capability to survey large and complete (magnitude
limited) samples of objects; and (3) the matching collection of synoptic and multi-epoch
spectrophotometric and radial velocity measurements \citep[cf.][]{Lindegren2008}. The range of science
questions that can be addressed with such a data set is immense and Gaia will revolutionize studies
of the Milky Way.  Moreover, such a massive survey is bound to uncover many surprises that the
universe still holds in store for us. The numerous science cases for Gaia can be found in, for
example, in the proceedings of the conferences `The Three-Dimensional Universe With Gaia'
\citep{Turon2005} and `Gaia: At the Frontiers of Astrometry' \citep{Turon2011}. A recent extensive
description of the Gaia mission and its expected performances was provided by \cite{DeBruijne2012}.

The astrometric measurements are collected employing a wide photometric band (the Gaia $G$ band)
which covers the range $330$--$1000$ nm. Multi-colour photometry will be obtained for all objects by
means of low-resolution spectrophotometry. The photometric instrument consists of two prisms
dispersing all the light entering the field of view. One disperser --- called BP for Blue Photometer
--- operates in the wavelength range $330$--$680$ nm; the other --- called RP for Red Photometer ---
covers the wavelength range $640$--$1000$ nm. In addition radial velocities with a precision of
$1$--$15$ km~s$^{-1}$ will be measured for all objects to $17^\mathrm{th}$ magnitude, thus
complementing the astrometry to provide full six-dimensional phase space information for the
brighter sources. The radial velocity instrument (RVS) is a near-infrared ($847$--$874$ nm,
$\lambda/\Delta\lambda\sim 11\,000$) integral-field spectrograph dispersing all the light entering
the field of view.

\begin{figure}[t]
  \centering
  \includegraphics[width=0.97\textwidth]{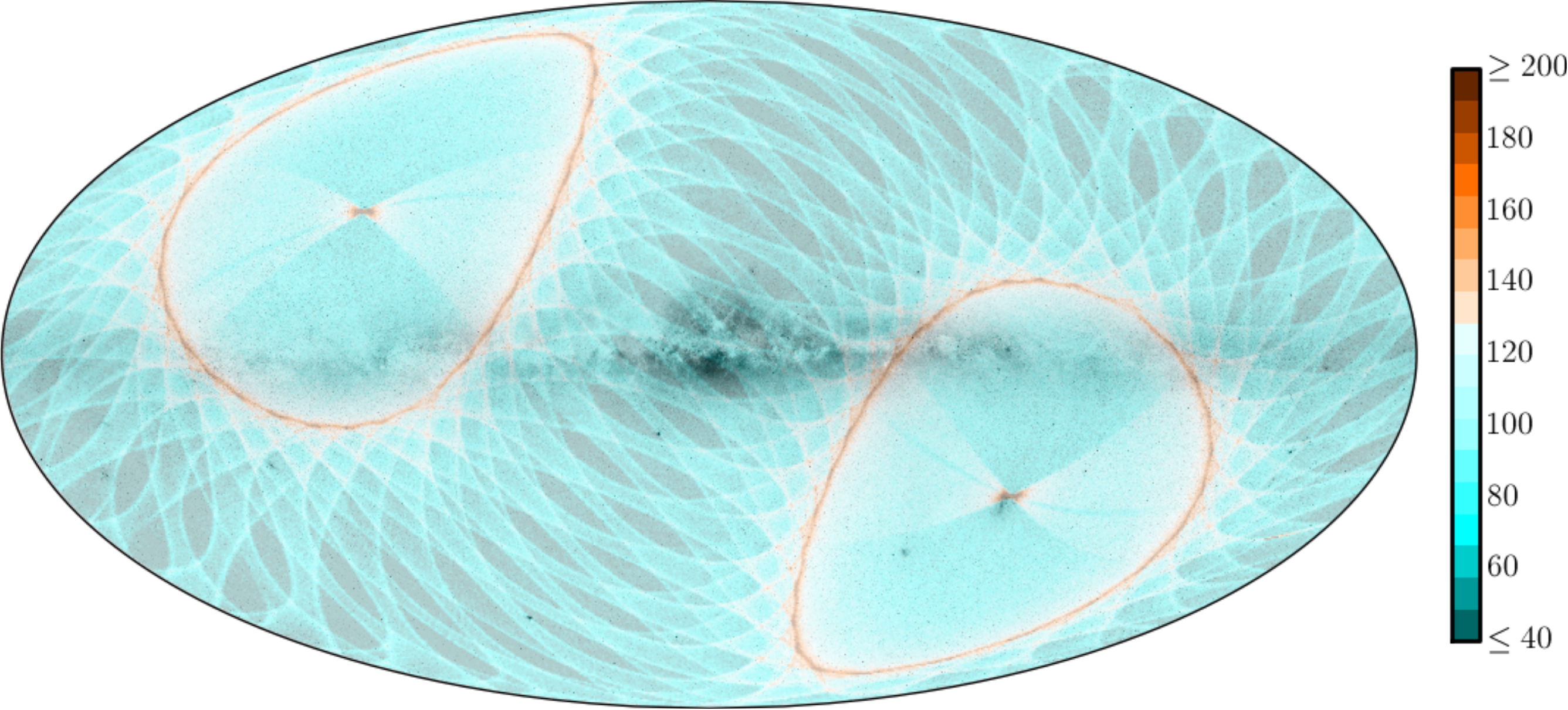}
  \caption{The predicted sky coverage of Gaia in Galactic coordinates projected on an image of the
    night sky. The colours indicate the number of times a particular region on the sky is
    observed. The two annuli where most of the observations are reflect the constant $45^\circ$
    angle between the Gaia spin axis and the direction from the spacecraft to the sun. The extra
    coverage around the ecliptic poles reflects the early phases of the mission when these areas
    will be covered repeatedly. The scan law simulation was done with the DPAC AGISLab software
    \citep{Holl2010}. The background image is the red band of the ESO Milky Way panorama (ESO/S.\
    Brunier).\label{fig:scanning}}
\end{figure}

The focal plane of Gaia comprises an array of 106 CCD detectors which serve the three instruments
mentioned above. The observing programme is based on the autonomous on-board detection of celestial
sources, which is unbiased and complete to $G=20$ ($V\sim20$--$22$). Gaia will be located at L2 and
scan the sky with its two telescopes by continuously spinning around the axis perpendicular to the
two lines of sight. The spin axis in addition makes a precessing motion around the spacecraft-sun
direction, and as a result Gaia will scan the whole sky roughly every 6 months. Each celestial
source will on average be observed about 70 times during the 5 year mission lifetime with a
quasi-regular time sampling. The number of times a source is observed is not uniform across the sky
as illustrated in figure \ref{fig:scanning}. The regions in the annuli located 45$^\circ$ away from
the ecliptic poles are observed most often, while the regions around the ecliptic are covered less
often. The number of stars in the Gaia catalogue is estimated to be $\sim 7\times10^5$ to $G=10$,
$48\times10^6$ to $G=15$ and $1.1\times10^9$ to $G=20$. About $60$ million stars are expected to be
seen as binary or multiple systems by Gaia, among which about $10^6$--$10^7$ eclipsing binaries. In
addition the catalogue will contain astrometry and photometry for $\sim3\times10^5$ solar systems
bodies, $\sim5\times 10^5$ quasars, and some $10^6$--$10^7$ galaxies. The sky survey by Gaia will
also produce the most accurate optical all-sky map ever, with an angular resolution comparable to
that of the Hubble Space Telescope.

\section{Scientific performances}

\begin{figure}[t]
  \centering
  \includegraphics[width=0.8\textwidth]{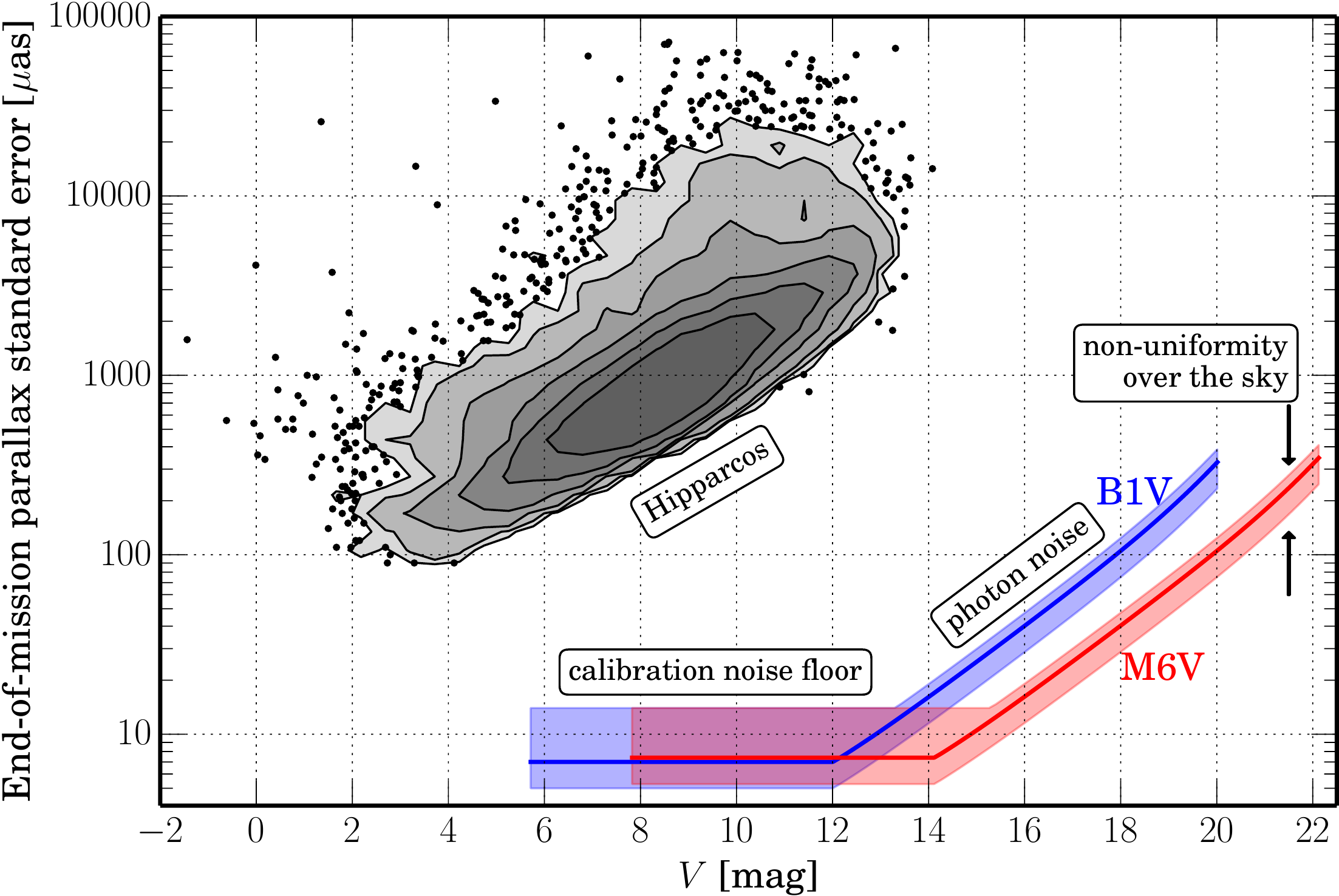}
  \caption{Parallax accuracies as a function of source brightness in the $V$-band for Gaia and
    Hipparcos.The contours and dots show the Hipparcos parallax errors, where the values are taken
    from \cite{FVL2007}. The lines show the predictions of the Gaia sky averaged parallax standard
    errors. At the bright end calibration errors will dominate and the parallax accuracies will
    range from $\sim 5$ to $\sim14$ $\mu$as. At the faint end the behaviour of errors as a function
    of $V$ is dictated by photon noise. The parallax accuracies are shown for an early and a
    late spectral type star to illustrate that at a given $V$ the astrometric accuracies are better
    for red stars. As a function of $G$ the differences are negligible. The bands around the average
    relations reflect the uncertain calibration errors at the bright end and the variation in sky
    coverage at the faint end. The accuracy predictions (obtained from the Gaia web pages) include a
    rough estimate of the effects of radiation damage and a 20\% margin (factor 1.2) to account for
    unmodelled effects. The standard errors in position and proper motion can be obtained by
    applying factors of $\sim0.7$ and $\sim0.5$, respectively, to the parallax standard
    errors.\label{fig:hiptogaia}}
\end{figure}

The expected scientific performance of Gaia in terms of the astrometric, photometric, and radial
velocity accuracies achieved is described in \cite{DeBruijne2012} and more details can be found on
the Gaia web pages at the following link:
\url{http://www.rssd.esa.int/index.php?project=GAIA&page=Science_Performance}. The performance
predictions have been confirmed following extensive tests conducted in December 2012 with the Gaia
payload module in cold vacuum. Figure \ref{fig:hiptogaia} shows the dramatic improvement over the
Hipparcos mission, in both astrometric accuracy and survey depth, that will be achieved by Gaia. 

The photometric capabilities of Gaia are described in detail in \cite{Jordi2010}, while the expected
accuracies to which stars can be characterized (in terms of temperature, surface gravity,
metallicity, extinction) on the basis of the Gaia photometry can be found in \cite{Liu2012}. The RVS
instrument of Gaia and its science capabilities were most recently summarized by
\cite{CropperKatz2011} and \cite{KatzEtAl2011}. Much more detail on the RVS can be found in
\cite{KatzEtAl2004} and \cite{WilkinsonEtAl2005}.

\section{Launch and commissioning}

Gaia will be launched on November 20 2013 from the European spaceport in French Guiana by a
Soyuz-STB/Fregat rocket. After launch the spacecraft will be inserted into a trajectory to
L2 where it will arrive in a few weeks.
Soon after the commissioning of Gaia's scientific instruments will take place, an activity which is
expected to take about 3--4 months. During this phase Gaia will be operated in a special survey mode
which ensures that it repeatedly scans over the Ecliptic Pole regions which have been extensively
surveyed from the ground in anticipation of the mission. The data so collected will allow a detailed
check of the performances of the Gaia instruments in terms of the achieved source detection
efficiency, image quality, photometric throughput, spectroscopic resolving power, and attitude
noise. These ingredients will allow a detailed assessment of the expected Gaia astrometric,
photometric and spectroscopic (radial velocity) performances.

During this phase the Gaia Data Processing and Analysis Consortium (cf.\ section \ref{sec:dpac})
will process the Gaia telemetry and participate in the detailed performance verification for Gaia.
The assessment of the Gaia instruments as they actually behave in flight and updated science
performance predictions will be published in the course of 2014 in order to give the scientific
community an early insight into what can be expected from Gaia.

\section{Gaia data processing}
\label{sec:dpac}

The on-ground data processing for Gaia is a very large and highly complex task, linking all
astrometric, photometric and radial velocity measurements into a large iterative solution. For the
astrometric instrument the iterative solution is aimed at transforming the source image location
measurements in pixel coordinates to angular field coordinates through a geometrical calibration of
the focal plane, and subsequently to coordinates on the sky through calibrations of the instrument
attitude and the basic angle between the lines of sight of the two telescopes. Moreover, corrections
for systematic chromatic shifts need to be made (using the photometric measurements), as well as
aberration corrections, corrections for perspective acceleration (involving the RVS measurements)
and corrections for general-relativistic light bending due to the Sun, the major planets, some of
their moons, and the most massive asteroids. Image location shifts caused by the radiation damage
induced stochastic charge trapping and de-trapping in CCDs also need to be understood and calibrated
with high precision. Repeated observations by Gaia of every star permit a complete determination of
each star's five basic astrometric parameters --- two angular positions, two proper motions and the
parallax. More information on the astrometric processing can be found in \cite{LindegrenEtAl2012}.

The treatment of the spectrophotometry \cite[cf.][]{Busso2012} starts with the pre-processing of the
raw dispersed images (e.g., bias and sky background removal) and the disentangling of overlapping
images in crowded fields. This is followed by an iterative process of internal calibration in which
all measurements are transformed to the same mean instrumental system by accounting for differences
across the focal plane in the prism dispersion curves, point spread function, and geometric calibration,
as well as performing flat fielding and correcting for CTI effects. The last step is to perform and
absolute calibration of the spectrophotometry using standard stars.

For the radial velocity spectrograph similar processing steps are taken \citep{KatzEtAl2011}. In
addition the very low signal levels at the faint end of the RVS magnitude range mandate a careful
`stacking' of multiple spectra collected for each source in order to allow the determination of
the radial velocity. At the bright end detailed epoch spectra can be obtained for each source. The
RVS instrument is especially sensitive to the effects of CTI and electronic bias non-uniformities
which have to be modelled and accounted for in the calibrations.  The data processing for both the
RVS and photometric instruments relies on the knowledge of source positions on the sky and of the
spacecraft attitude, quantities that are derived in the astrometric data processing.

This highly interlinked `upstream' processing produces raw astrometric, photometric and RVS results
which are further processed. A detailed analysis will be made of multiple stars, extended sources,
galaxies, exoplanets and solar system objects and the spectra from the photometers and RVS will be
used to characterize the astrophysical properties of every source observed by Gaia. Finally the
repeated observations of each source can be used to carry out a detailed variability analysis.  More
details on these `downstream' processing tasks can be found in \cite{Pourbaix2011} (double and
multiple stars), \cite{Sozzetti2013} (exoplanets) \cite{Tanga2012} (solar system objects),
\cite{Tsalmantza2012, KroneMartins2013} (galaxies), \cite{BailerJonesEtAl2013} (astrophysical
parameters of Gaia sources), and \cite{Eyer2012} (variable stars).

\begin{figure}[t]
  \centering
  \includegraphics[width=0.9\textwidth]{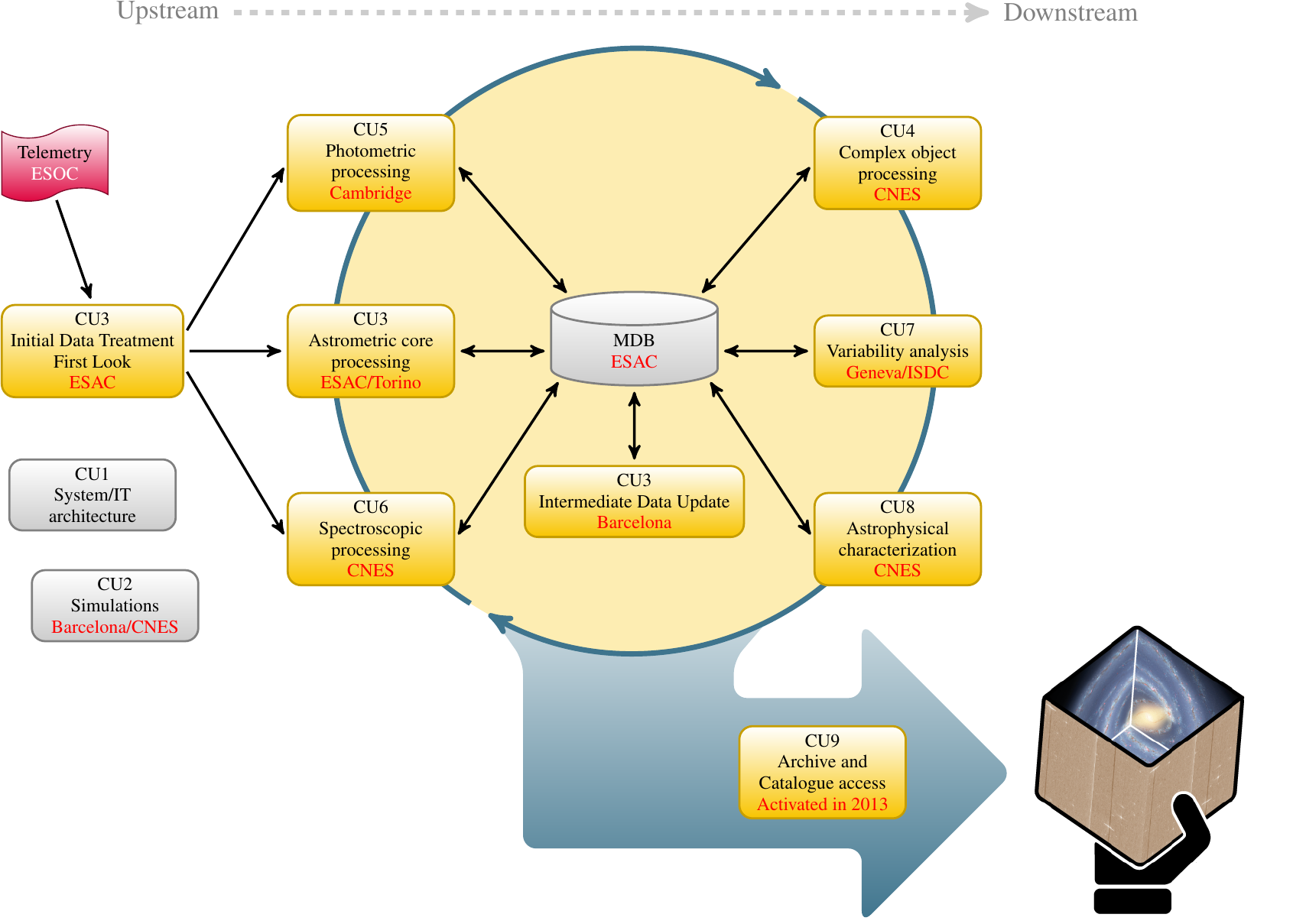}
  \caption{Schematic overview of the organization of the data flow withing the Gaia Data Processing
    and Analysis Consortium. For each DPAC coordination unit the processing tasks listed in the
    boxes are carried out at the data processing centre indicated in red. See text for more
    explanations.\label{fig:dpac}}
\end{figure}

The multitude of tasks described above will be undertaken by the scientific community in Europe
which has organized itself into the Gaia Data Processing and Analysis Consortium (DPAC). The data
processing activities will be structured around nine `coordination units' (CUs) and six data
processing centres. Each CU is responsible for delivering a specific part of the overall data
processing system for Gaia. The role of each CU and of the data processing centres is illustrated
schematically in figure \ref{fig:dpac}.

In this diagram the flow of data within DPAC and between the data processing centres is shown. The
Gaia telemetry is sent from ESOC to the Science Operations Centre at ESAC where it is unpacked and
processed by the Initial Data Treatment and First Look pipelines. The latter serves to monitor the
instrument health of Gaia in great detail while the former prepares the raw telemetry for further
processing by CUs 3 (astrometry), 5 (photometry) and 6 (spectroscopy and radial velocities). The
results from these `upstream' CUs are then stored in the Main Data Base (MDB) housed at ESAC. The
`downstream' CUs 4, 7, and 8 then pick up the processing results from the MDB and carry out the
processing of non-single stars and non-stellar sources, the classification and characterization of
all sources, and the variability analysis (detection, classification, and derivation of variable
source light curves). The results from these CUs also go back into the MDB. The iterative loop is
closed by the intermediate data update in which improved instrument calibrations are derived (taking
into account the results from all CUs 3/5/6 and 4/7/8) which are then applied to derive improved
upstream and downstream data products. Finally, when a particular version of the Main Data Base
contents is deemed of sufficient quality to warrant a data release the MDB contents are passed onto
CU9 for extensive validation, documentation, and release of the Gaia processing results.

\section{Data release scenario}

The final Gaia catalogue release is foreseen to take place around 2022. This is of course still some
time away and hence in the mean time intermediate data releases are foreseen. The data release
scenario for Gaia can be found on the Gaia web pages at
\url{http://www.rssd.esa.int/index.php?project=GAIA&page=Data_Releases}. The release scenario has
been designed by carefully considering the complex data processing described above, the available
staff effort within DPAC, and assuming smooth operations from the start. Hence, the precise times of
data releases cannot be fixed at this moment.

The very first data to be released from the Gaia mission will be the data associated with the
so-called Science and Solar System Alerts streams. The Science Alerts concern transient sources that
should be followed up promptly from the ground, while the solar system alerts are intended for
follow-up observations of newly discovered or very fast moving solar system objects.

The first catalogue release is foreseen to take place 22 months after launch (so roughly end 2015)
and will consist of an all sky map (positions and magnitudes) for most of the Gaia sources. In
addition, for stars in common with the Hipparcos Catalogue, Gaia and Hipparcos positions will be
combined to give very accurate ($\sim 30$--$190$~$\mu$as/yr) proper motions
\citep[see][]{DeBruijneEilers2012}.

The list of subsequently planned data releases can be found on the pages above and will include data
of increasing accuracy and diversity. It should be stressed here that none of the data releases will
be preceded by a proprietary period for the DPAC. The releases are immediately publicly available
world wide. This is in fact a unique aspect of the Gaia mission. More information on the expected
Gaia catalogue contents and the ideas being developed for making the data available can be found in
\cite{Brown2012} and \cite{LuriEtAl2013}.\footnote{Available at
\url{http://www.rssd.esa.int/SA/GAIA/docs/library/XL-033.htm}.}

\section{Preparing yourself for Gaia}

The Gaia mission will result in a large catalogue (over 1 billion sources) containing a very rich
diversity of information. Scientifically exploiting such a data set will not be trivial and thus
some preparation within the astronomical community is required. The Gaia community will support this
process and I give here some examples of activities that astronomers can profit from in their
preparations for the use of the Gaia data.
\begin{itemize}
  \item The Gaia Science Performance web
    pages\footnote{\url{http://www.rssd.esa.int/index.php?project=GAIA&page=Science_Performance}}
    provide, among others, background information on the instruments and the error modelling,
    interpolation tables and formulae for simulating the errors, the predicted variations of errors
    over the sky, transformations from the Johnson and Sloan systems to Gaia photometric system, and
    references to the relevant literature. The information on these pages allows one to
    realistically simulate the performance of Gaia in order to prepare for the scientific
    exploitation of the data.
  \item Simulated Gaia catalogues are being created by DPAC and are made available through, for
    example, the Centre de Donn\'ees astronomiques de Strasbourg (CDS). The Gaia Universe Model
    Snapshot has already been made available \citep[see][]{Robin2012}. It contains a simulated
    `universe' (i.e.\ true properties of sources observed by Gaia) from which the simulated Gaia
    catalogue is subsequently generated (observable quantities and their errors). The simulated
    catalogue is currently in production and will be made available later this year. These simulated
    data sets are representative of the actual Gaia catalogue and can be used to exercise one's data
    analysis algorithms.
  \item Since 2009 the Gaia Research for European Astronomy Training (GREAT)
    network\footnote{\url{http://www.ast.cam.ac.uk/GREAT}}, funded through the European Science
    Foundation, has been stimulating activities in anticipation of the Gaia mission. Numerous
    workshops and conferences have been organized, focusing on Gaia science topics and how to
    prepare for exploiting the mission data. The topical working groups within the network are still
    open to new members, contributions, and ideas. Refer to the GREAT web pages for more
    information. A notable outcome of the GREAT networking activities is the Gaia-ESO survey
    \citep{GaiaESO2012}.
  \item The Gaia community is now in the process of defining how the Gaia results will be made
    accessible to astronomers. This includes developing the data archiving and querying systems, as
    well as producing detailed documentation and providing sophisticated data analysis  and data
    mining tools \citep{LuriEtAl2013}. The astronomical community was consulted, through the GREAT
    network, to collect requirements on the way the Gaia data should be made available. This was
    done by inviting the submission of `use cases' which will drive the requirements specifications
    for the Gaia catalogue and archive access mechanisms. Your ideas are still welcome and can be
    submitted through the following wiki pages:
    \url{http://great.ast.cam.ac.uk/Greatwiki/GaiaDataAccess}.
\end{itemize}


\bibliographystyle{aa}  
\bibliography{brown} 

\end{document}